\documentclass[aps,prl,twocolumn,showpacs,letterpaper,superscriptaddress]{revtex4-1}

\usepackage[colorlinks=true, citecolor=blue, linkcolor=blue, urlcolor=blue]{hyperref}

\usepackage{graphicx,dcolumn,longtable,epsfig}
\usepackage[usenames]{color}
\usepackage{amssymb}
\usepackage{amsmath}
\usepackage{epstopdf}
\usepackage{bm}
\usepackage{footnote}
\usepackage{float}
\usepackage{subfigure}
\usepackage{color}
\usepackage{ulem}

\input epsf.tex

\def\bea{\begin{eqnarray}}
\def\eea{\end{eqnarray}}
\def\be{\begin{equation}}
\def\ee{\end{equation}}

\begin{document}
\title{A quantum Monte Carlo study of the long-ranged site-diluted XXZ-model as realized by polar molecules}
\author{ C. Zhang }
\affiliation{Homer L. Dodge Department of Physics and Astronomy,
The University of Oklahoma, Norman, Oklahoma ,73019, USA}
\author{B. Capogrosso-Sansone}
\affiliation{Department of Physics,
Clark University, Worcester, Massachusetts, 01610, USA}

\begin{abstract}
Motivated by recent experiments with ultracold polar molecules trapped in deep optical lattices, we study ground-state properties of the site-diluted long-ranged XXZ model. Site dilution results in off-diagonal disorder.
We map the spin model to a hard-core Bose-Hubbard model and perform large-scale Monte Carlo simulations by the Worm algorithm. In absence of site-dilution, we find that, 
for large enough interaction, three phases are stabilized: a superfluid phase, a checkerboard solid phase, only present at density $m = 0.5$, and a checkerboard supersolid phase which can be reached by doping the CB phase away from half-filling. In the presence site-dilution and at fixed density $m=0.5$, we find that, unlike what observed in the case of short-range hopping, localization never occurs even for site dilution larger than the percolation threshold, and off-diagonal order, though strongly suppressed, persists for arbitrarily large values of site-dilution. 
 
\end{abstract}

\maketitle

\section{Introduction}
Interacting bosons in the presence of disorder have attracted a great deal of attention in the past decades. Disorder can be found or engineered in a variety of systems, ranging from $^4He$ in porous media and aerogels~\cite{Crowell:1995cx, Crowell:1997gm, Reppy:2000cb}, thin superconducting films \cite{Goldman:1998jy, Vitkalov:2001in, Schneider:2012id}, Josephson-junction arrays~\cite{vanderZant:1996cf} to ultracold gases~\cite{DeMarco2009PRL, DeMarco2010Nature, Inguscio2008}. Diagonal disorder has been extensively studied. The compressible gapless Bose glass phase, intervening between the Mott-insulator and the superfluid phase in the presence of diagonal disorder, has been investigated in great detail using a variety of different methods such as density matrix renormalization group~\cite{Rapsch:1999bw}, Monte Carlo simulations~\cite{Ceperley1991, Pollet2009BG, Pollet2009PRB, Soyler:2011ik, Ceperley2011, Zhang:2015it}, and mean field theory~\cite{Niederle:2013jy}. Diagonal disorder has also been studied in the presence of long-range hopping. Depending on the ratio between the power-law decay of the hopping and the dimensionality of the system, the combination of on-site disorder and long-range hopping may result in localization, critical behavior or fully extended states~\cite{Levitov:1990jd}. On the other hand, systems exhibiting off-diagonal disorder have received less attention. As an example, it has been shown that in the presence of off-diagonal disorder, the incompressible yet gapless Mott glass phase  intervenes between the Mott-insulator and the superfluid phase~\cite{Sengupta:2007jx, Prokofev:2004fk, Roscilde:2007ch, Vojta:2016kl}.   
When both, off-diagonal long-range interaction and off-diagonal disorder are present, the interplay between the two may result in new interesting phenomena.

Purely off-diagonal disorder in systems exhibiting long-range hopping and interactions have recently been realized experimentally with polar molecules trapped in deep optical lattices~\cite{Yan:2013fn, Hazzard:2014bx}. A spin 1/2 degree of freedom is encoded in two internal states of the molecules (the lowest rovibrational state and an excited rotational state) which are coupled via a microwave field. Molecules are pinned to lattice sites due to the deep optical lattice while the dipolar interaction induces spin exchanges between pairs of molecules. At integer unit filling, and in the presence of an external DC electric field, this system realizes a spin XXZ model where the effective magnetic interactions decay as $\frac{1}{r^3}$ where $r$ is the distance between lattice sites~\cite{Hazzard:2014bx}. Spin-spin interactions can be tuned with the external electric field but they can also be modified by choosing different pairs of rotational levels. In a typical experiment, not all sites are occupied, but rather, sites are randomly filled by a single molecule or are unoccupied. Recent experiments report a filling fraction of $~25\%$~\cite{Moses:2015df}. This results in the presence of disorder in the long-range spin-exchange term (equivalent to long-range hopping). Since the configuration of occupied sites varies in the experiments from shot to shot, experimental measurements are effectively averaging over different disorder realizations. Recently, this system has been studied in the case of a single spin-excitation present~\cite{Deng:2017uu, Deng:2016gs}. The authors found that localization of eigenstates depends on dimensionality and filling. While these results can be extended to the case of a dilute gas of excitations, the full many-body case has yet to be studied. Here, we study the site-diluted long-range XXZ model. Site dilution results in off-diagonal disorder. We find that, unlike what observed in the case of short-range interactions, localization never occurs even for site-dilution larger than the percolation threshold and off-diagonal order, though strongly suppressed, persists for arbitrarily large values of site-dilution.

\section{Hamiltonian}
In the following we study the site-diluted two-dimensional spin 1/2 XXZ model defined on a square lattice of linear size L, as realized by polar molecules trapped in deep optical lattices. The spin 1/2 degree of freedom is encoded in two rotational states coupled via a microwave field, and the quantization axis defining the two rotational states is aligned perpendicular to the 2D plane of the square lattice. Hopping is suppressed and at most a single molecule occupy each lattice site. The model reads as:

\begin{equation}
\label{eq:H1}
H=\sum_{i,j}\Big{[}\frac{J_{ij}^{\perp}}{2} (S_{i}^{+}S_{j}^{-}+S_{i}^{-}S_{j}^{+})+J_{ij}^{z}(S_{i}^z S_{j}^z) \Big{]}
\end{equation}

Here $S_{i}^{\pm}$ and $S^z$ are the spin 1/2 operators which obey $[S_{i}^{z}, S_{j}^{\pm}]=\pm \delta_{ij}S_{i}^{\pm}$; the spin-exchange interaction $J_{ij}^{\perp}=\frac{J_{\perp}}{r_{ij}^3}$ with $J_{\perp}<0$;
$J_{ij}^z=\frac{J_{z}}{r_{ij}^3}$  is the strength of the spin-spin repulsion; $r_{ij}$ is the relative distance between site $i$ and site $j$.
Note that the Ising term, with coupling strength $J_z$, is present only if an external DC electric field (parallel to the quantization axis) is applied. Model~\ref{eq:H1} features purely off-diagonal disorder resulting from randomly distributed unoccupied lattice sites.

Model~\ref{eq:H1} can be exactly mapped onto a site-diluted model of hardcore lattice bosons in the presence of long-range dipolar hopping and dipolar interaction:

\begin{align}
\label{eq:H}
\nonumber H&=-J\sum_{ i<j }\frac{1}{r_{ij}^3}(a_i^\dagger a_j+a_i a_j^\dagger ) \\
&+V\sum_{i<j}{\frac{n_{i}n_{j}}{r_{ij}^{3}}}
\end{align}
where $\frac{J_{\perp}}{2}=-J$, $J_z=V$
. The first term in the Hamiltonian is the kinetic energy, where $a_i^\dagger$ ($a_i$) are the bosonic creation (annihilation) operators on the site $i$ satisfying usual commutation relations and the hard-core constraint $a_i^\dagger a_i^\dagger=0$. $J$ is the hopping amplitude with hopping matrix element $J_{ij}=\frac{J}{r_{ij}^3}$, where $r_{ij}$ is the relative distance between site $i$ and site $j$. 
The second term describes the dipole-dipole purely repulsive interaction with strength $V$ and $V_{ij}=\frac{V}{r_{ij}^3}$. 
Site-dilution corresponds to removing a certain fraction of sites from the lattice, which, in the experimental setup, correspond to unoccupied sites. Site-dilution results in off-diagonal disorder.

In the following we perform large-scale quantum Monte Carlo simulations by the Worm algorithm~\cite{Prokofev:1998gz} to study equilibrium phases of model~\ref{eq:H} both without and with site-dilution.

\section{Phase diagram in absence of site-dilution}
\begin{figure}[h]
\includegraphics[trim=0cm 1.5cm 0cm 2.5cm, clip=true, width=0.5\textwidth]{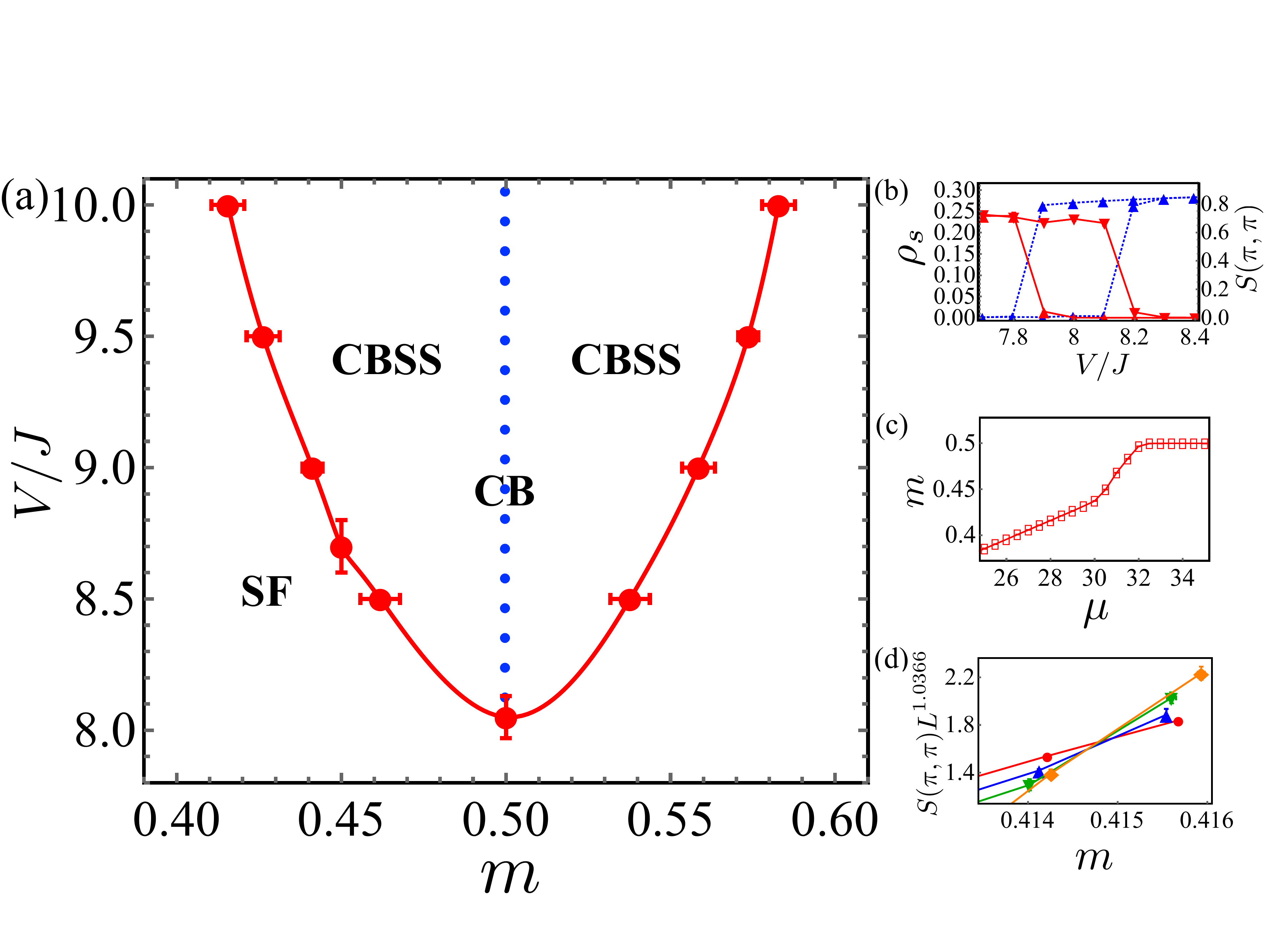}
\caption{(Color online) (a) Ground state phase diagram of model~\ref{eq:H} as a function of interaction strength V/J and filling factor $m$. Error bars come from finite size scaling. The system features three phases: a superfluid phase (SF), a checkerboard solid phase (CB) and a checkerboard supersolid phase (CB SS). At half filling, the SF is destroyed in favor of a CB via a first order phase transition. At $V/J>8$, the CB SS can be reached by doping the CB solid with particles or holes. At large enough doping, the CB SS gives way to a SF via a second order phase transition. Red circles are numerical results for the SF to CB SS second order transition. Blue dotted line indicates CB at half filing. (b) hysteresis curves for $\rho_{s}$ (red down triangles) and $S(\pi, \pi)$ (blue up triangles) as a function of $V/J$ for system size $L=32$ signaling a first-order CB-SF phase transition at $m=0.5$. (c) density $m$ as a function of the chemical potential $\mu$ for $V/J=9.0$. (d) finite size scaling of $S(\mathbf{\pi}, \mathbf{\pi})$ at fixed $V/J=10.0$ for system size $L=20$, 24, 28 and 32 (red circles, blue up triangles, green down triangles and orange diamonds, respectively). The crossing of different curves marks the transition point at $m=0.4147 \pm 0.005$. Error bars in (b), (c), (d) come from statistical Monte Carlo sampling. Here and throughout the text, when not visible, error bars are within symbol size.} 
\label{FIG1}
\end{figure}

In this Section we study the phase diagram of model~\ref{eq:H} in the absence of site-dilution.
The phase diagram is shown in Figure~\ref{FIG1} (a) in the plane of $V/J$ vs. filling factor $m=\langle S_z \rangle$. 
We find that, for large enough interaction, three phases are stabilized:  a superfluid phase (SF), characterized by off-diagonal long-range order, a checkerboard solid phase (CB), characterized by diagonal long-range order, only present at density $m=0.5$, and a checkerboard supersolid phase (CB SS), characterized by both diagonal and off-diagonal long-range order, which can be reached by doping the CB phase away from half-filling. On the other hand, at lower interaction strength, the system is in a SF phase for any value of the filling factor. In spin language, the CB order corresponds to the easy-axis antiferromagnetic order, the SF phase corresponds to easy-plane ferromagnetic order.
The SF phase possesses finite superfluid density $\rho_{s}$, easily accessible in our simulations. Notice that, since the hopping term in Eq.~\ref{eq:H} is not limited to nearest neighbors, the standard expression of superfluid density in terms of winding numbers 
~\cite{Winding} must be generalized~\cite{Rousseau:2014jt}. The CB solid is characterized by a finite value of  structure factor $S(\mathbf{k})=\sum_{\mathbf{r},\mathbf{r'}} \exp{[i \mathbf{k} (\mathbf{r}-\mathbf{r'})]\langle n_{\mathbf{r}}n_{\mathbf{r'}}\rangle}/N$, $\mathbf{k}$ is the reciprocal lattice vector with $\mathbf{k}=(\pi, \pi)$ in the CB. Finally the CB SS is characterized by finite $\rho_{s}$ and $S(\pi, \pi)$.
Compared to the case of nearest-neighbor hopping only, in the presence of long-range hopping, the SF phase is much more robust against interaction strength. Indeed, at $m=0.5$, the CB phase (blue dotted line in Figure~\ref{FIG1} (a) ) is reached for interaction $V/J\sim 8$ compared to $V/J\sim 3.5$~\cite{CapogrossoSansone:2010em} in the nearest-neighbor hopping case. Because the presence of long-range hopping breaks down the sublattice invariance, model~(\ref{eq:H}) does not have a Heisenberg point (i.e. $J_x=J_y=J_z$) and so the transition from SF to CB does not happen continuously via the Heisenberg point but, rather, it is expected to be of first order since it is a transition between two broken-symmetry states.  Our simulations results confirm this expectation as shown in Figure~\ref{FIG1} (b) where we show the hysteresis curve for $\rho_{s}$ (red down triangles) and $S(\pi, \pi)$ (blue up triangles)  as a function of $V/J$ for system size $L=32$. 

At fixed interaction strength $V/J>8$, one would not expect two order parameters to disappear continuously at the same critical point, hence, either the CB-SF transition is of first order or it happens via an intermediate phase where both orders coexist.
We have observed a stable CB SS as demonstrated by the lack of hysteresis in the density $m$ vs. chemical potential $\mu$ at fixed $V/J$. Moreover, as shown in Figure~\ref{FIG1} (c) for $V/J=9.0$, the $m$ vs $\mu$ curve does not display any discontinuity. The CB SS can be reached by doping the CB solid with particles or holes (particle-hole symmetry exists since bosons are hardcore). 
 At large enough doping, the diagonal order becomes unstable and the CB SS disappears in favor of a SF via a second order phase transition belonging to the (2+1)D Ising universality class. Standard finite size scaling is used to extract  transition points (red circles in Figure~\ref{FIG1} (a)) as shown in Figure~\ref{FIG1} (d) where we plot $S(\mathbf{\pi}, \mathbf{\pi}) L^{1.0366}$ vs filling factor $m$ at fixed interaction strength $V/J=10.0$ for system sizes $L=20$, 24, 28 and 32. The crossing of curves corresponding to different system sizes signals the transition, in this case at $m=0.4147\pm 0.005$. 



We conclude this Section by considering the robustness of the quantum phases observed against thermal fluctuations. We have performed finite-temperature simulations at fixed $V/J=9.0$. The critical temperature for SF-to-normal Kosterlitz-Thouless  transition and CB-to-normal 2D Ising-type transition at densities $m=0.4$, $0.47$ and $0.5$, corresponding to SF, CB SS, and CB phase at T=0 respectively, are all of the order of $T_{c}/J\sim 2$. For $J=104 h$ Hz and $V=0$ as in~\cite{Yan:2013fn}, this translates to a normal-to-SF transition temperature of about 10.0 nK.



\section{ SITE-DILUTED XXZ MODEL }
In this Section we study model~\ref{eq:H} in the presence of site-dilution. Because hopping is long-ranged, arguments based on percolation theory (see e.g.~\cite{Vojta:2016kl}) are inapplicable and the presence of an insulating phase beyond the percolation threshold is no longer guaranteed.
In the following, we fix the density of the system at $m=0.5$ (calculated with respect to available sites) and investigate localization of particles upon increasing site-dilution $p$ at given values of the interaction $V/J=0$, 2 and 7, which all correspond to a SF phase in the clean system. In all cases, temperature has been chosen in order to ensure that the system is in its ground state, i.e. $\beta=\frac{5}{6}L$ with L= 84, 100 and 150. As a reminder, site-dilution results from unoccupied sites in the experimental setups. Averaging over different realizations of site-dilution corresponds to averaging results from different experimental shots. The simulation results shown below are based on 50-100 realizations of site-dilution, depending on the value of the interaction and site-dilution. In order to see whether superfluidity is destroyed at large enough site-dilution, we look at the spatial decay of the off-diagonal correlator $f_{ij}\propto \sum_{\tau,\tau'}\langle a_{i}(\tau) a_{j}^\dag(\tau') \rangle$, where $a_{i}(\tau),\; a_{j}^\dag(\tau')$ are annihilation and creation bosonic operators expressed in the interaction picture. In the SF phase, $f_{ij}$ is expected to be long-ranged with respect to the distance $r_{ij}$ between sites $i$ and $j$.  

\begin{figure}[h]
\includegraphics[trim=0.6cm 1.5cm 1cm 3cm, clip=true, width=0.46\textwidth]{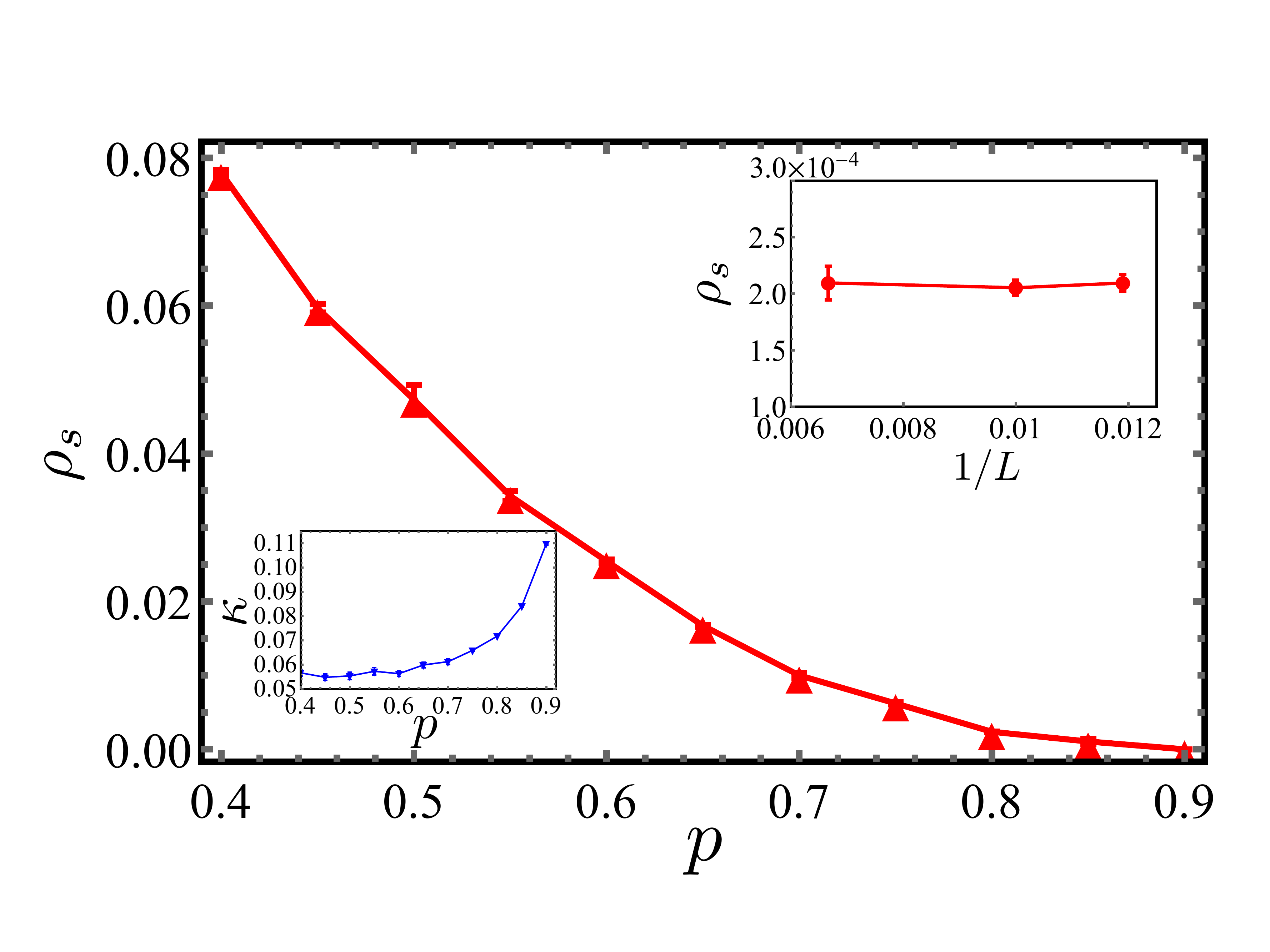}
\caption{(Color online) $V/J=0$. Main plot, superfluid density $\rho_s$ as a function of site-dilution $p$ for $L=84$. Superfluidity is strongly suppressed at larger site-dilution. Bottom-left inset, compressibility $\kappa$ as a function of site-dilution $p$. Compressibility increases at large enough site dilution. Top-right inset, superfluid density $\rho_s$ as a function of $1/L$ at site-dilution $p=0.9$. Error bars are a result of averaging over different site-dilution realizations.}
\label{FIG2}
\end{figure}
\begin{figure}[h]
\includegraphics[trim=0.5cm 2.1cm 0cm 2.5cm, clip=true, width=0.48\textwidth]{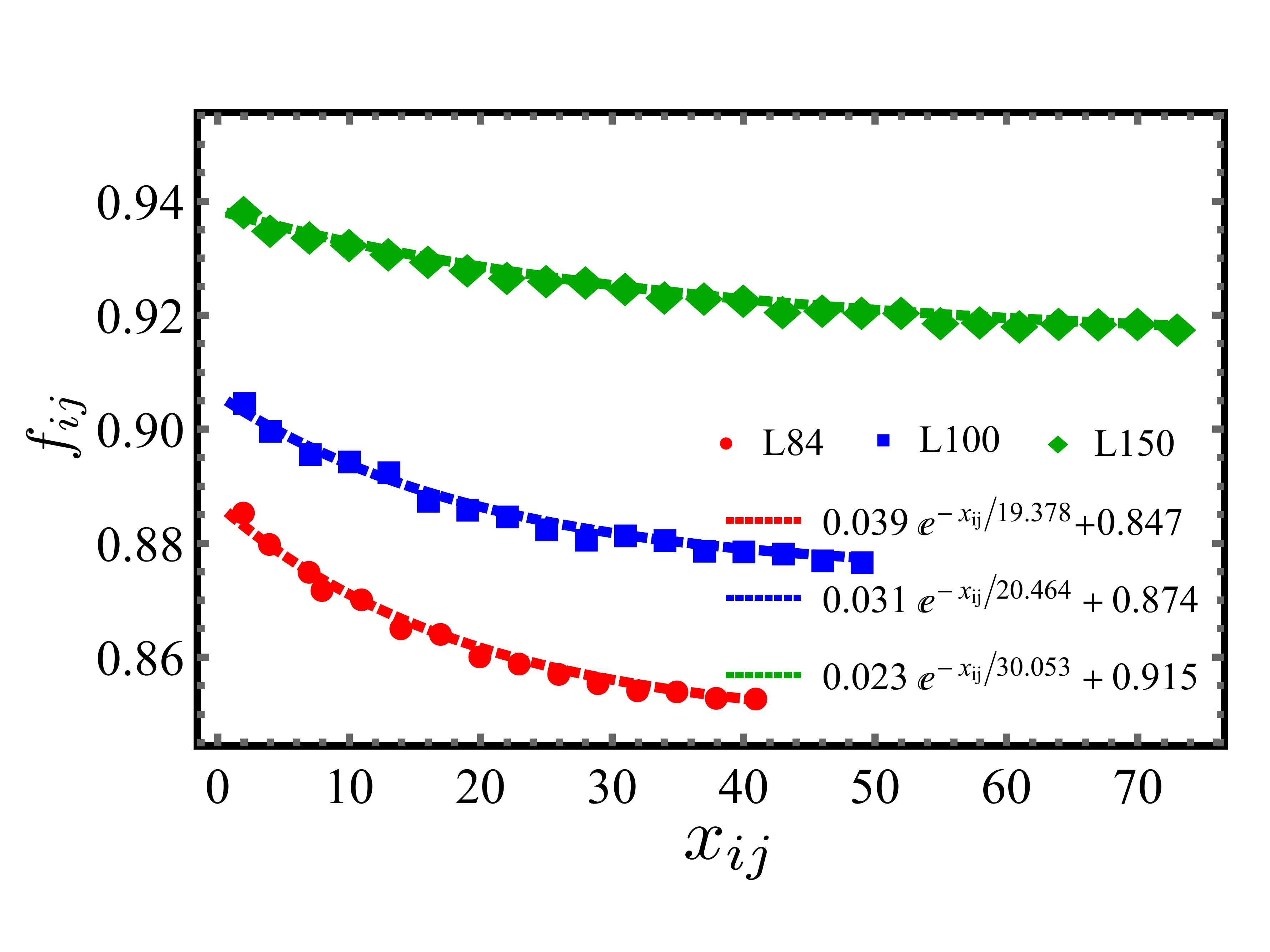}
\caption{(Color online) $V/J=0$, $p=0.9$ Correlation function $f_{ij}$ as a function of the $x$-distance between sites $i$ and $j$ for system size $L=84$, 100, and 150 (red circles, blue squares, green diamonds, respectively). Dotted lines are the exponential fit $a*e^{-x/\xi}+b$. The correlation function decays to a constant value different than zero, suggesting the system is in a SF phase. Note also that the correlation function $\xi$ increases with system size. Error bars are within symbol size and are a result of averaging over different site-dilution realizations.}
\label{FIG3}
\end{figure}

Let us start by considering the case $V/J=0$ where no diagonal interaction is present. This corresponds to absence of external DC electric field in an experimental setup. Figure~\ref{FIG2} shows superfluid density $\rho_s$ as a function of site dilution $p$ for system size $L=84$. As expected, superfluidity is suppressed as site-dilution increases. We have considered site dilution up to $p=0.9$. Even at such large dilution value, superfluidity, though strongly suppressed, still persists as demonstrated in the top-right inset of Figure~\ref{FIG2}, where we plot $\rho_s$ as a function of $1/L$ for $L=84$, $100$, $150$ and $p=0.9$ ($\rho_s$ saturates to a finite value with increasing $L$). To further support this conclusion, in Figure~\ref{FIG3} we plot $f_{ij}$ as a function of the distance along the $x$-direction for system sizes $L=84$, 100, and 150 and at fixed $p=0.9$. The correlation function exponentially decays to a constant value different than zero. Dotted lines in the figure are the result of the exponential fit $a*e^{-x/\xi}+b$ with $b\ne 0$. Note that, the correlation length $\xi$ increases with system size. These results strongly indicate that the system remains SF for arbitrarily large values of site dilution. This fact can be understood as follows. At site dilution $p$ larger than the percolation threshold $p_c=0.407253$~\cite{Vojta:2016kl}, SF islands within the lattice are geometrically disconnected but coherent! Superfluidity of the islands follows from the fact at $V/J=0$ the corresponding clean system is a SF.  Coherence between islands is guaranteed by long-range hopping. 
As site dilution increases, the average distance between SF islands also increases while hopping is further suppressed thus suppressing coherence and superfluidity. We notice that, a suppressed SF density will result in a suppressed critical temperature for the SF to normal liquid transition at finite temperature making it challenging to observe coherence experimentally (see also below).

\begin{figure}[h]
\includegraphics[trim=0cm 4cm 0cm 7cm, clip=true, width=0.45\textwidth]{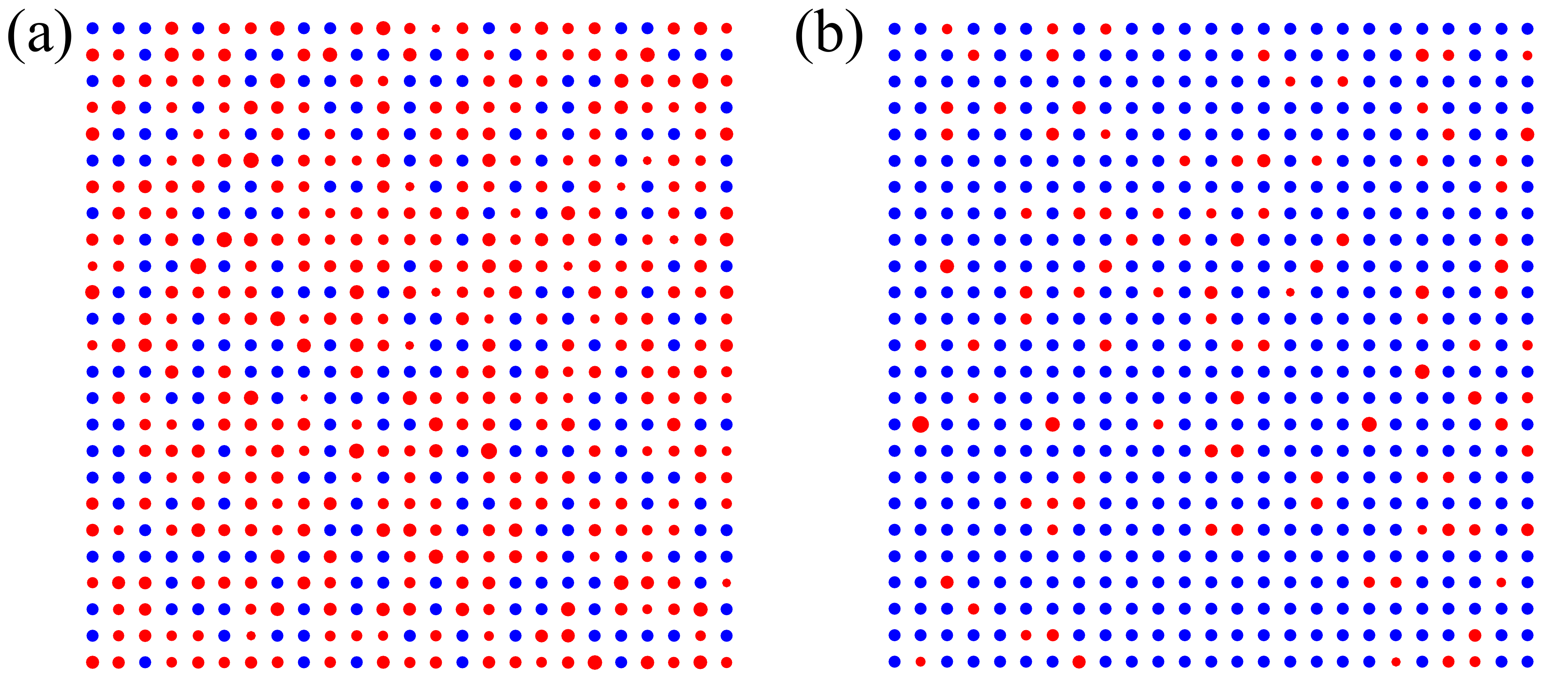}
\caption{(Color online) $V/J=0$. Imaginary time average of the density distribution within the lattice for a single Monte Carlo configuration and a single dilution realization, at site dilution $p=0.4$ (a) and $p=0.85$ (b). The radius of a red circle at a given site is proportional to the density at that site. Blue dots represent sites which have been removed from the lattice.}
\label{FIG4}
\end{figure}

Interestingly, as site dilution increases, compressibility also increases, as the bottom-left inset of Figure~\ref{FIG2} shows. Indeed, increased site-dilution corresponds to a larger average distance between disconnected lattice regions, making it harder for particles to delocalize within the entire lattice. This implies a non-uniform density distribution within the lattice and the presence of lower-density regions contributing to an increased compressibility. This can be seen in Figure~\ref{FIG4} where we show the imaginary time average of the density distribution within the lattice for a single Monte Carlo configuration and a single dilution realization, at site dilution $p=0.4$ (a) and $p=0.85$ (b). The radius of a red circle at a given site is proportional to the density at that site. Blue dots represent sites which have been removed from the lattice. The standard deviation of the average density $m=0.5$ corresponding to $p=0.4$ and $p=0.85$ are of the order of $0.045$ and $0.1$ respectively. The conspicuous presence of sites with significant lower than average density at $p=0.85$ is responsible for a larger compressibility. This scenario persists for configurations corresponding to larger and larger number of Monte Carlo steps. Once particles are able to reach a lower density region, they are `stuck'  there due to suppressed hopping and the density distribution remains non-uniform. 

\begin{figure}[h]
\includegraphics[trim=0.4cm 2cm 1cm 3cm, clip=true, width=0.48\textwidth]{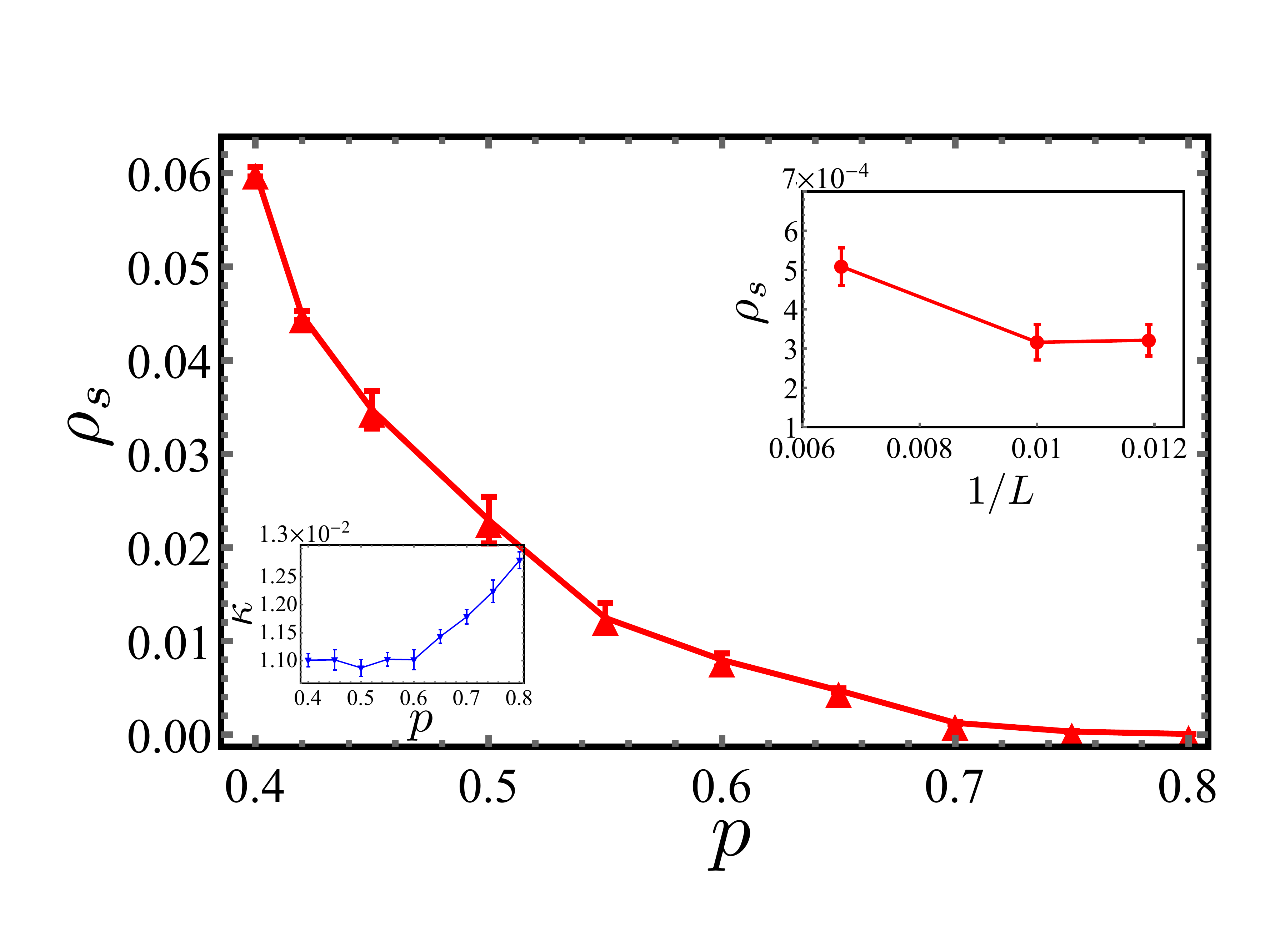}
\caption{(Color online) Same as Figure~\ref{FIG2} but for $V/J=7$ and $p=0.75$.
}
\label{FIG5}
\end{figure}

\begin{figure}[h]
\includegraphics[trim=0.5cm 2.2cm 0cm 2.5cm, clip=true, width=0.5\textwidth]{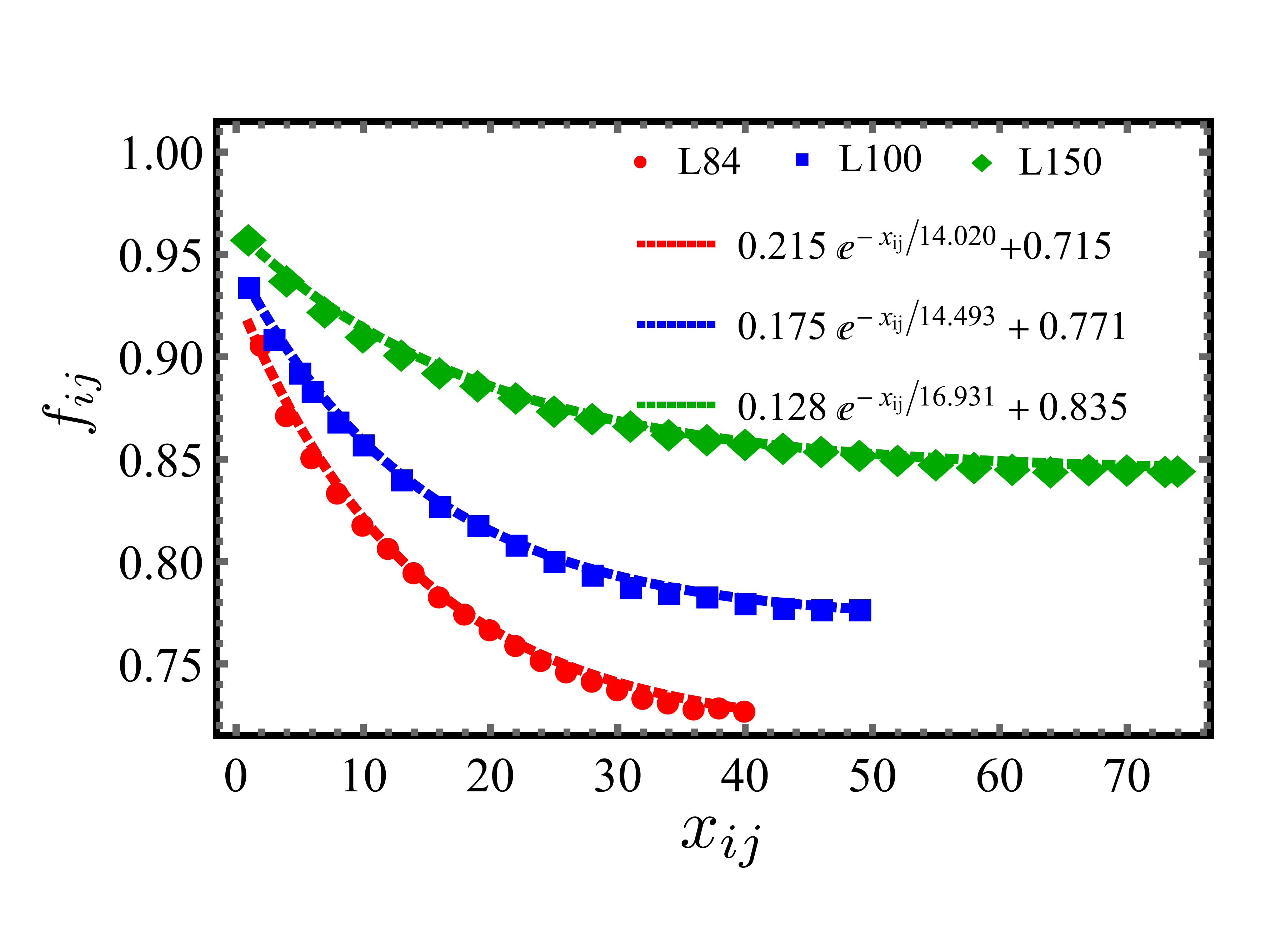}
\caption{(Color online) 
Same as Figure~\ref{FIG3} but for $V/J=7$ and $p=0.75$.
}
\label{FIG6}
\end{figure}

At fixed interaction strength $V/J=2$ and 7 we find qualitative similar results as for $V/J=0$. Here, we only report results corresponding to $V/J=7$. Figure~\ref{FIG5} shows superfluid density $\rho_s$ as a function of site dilution $p$ for $L=84$. Superfluidity, though suppressed, remains finite even for large site dilution. The top-right insert of Figure~\ref{FIG5} shows $\rho_s$ increasing with system size at fixed $p=0.75$, indicating that the system is SF. Figure~\ref{FIG6} shows the decay of the correlator $f_{ij}$ at $p=0.75$ for system sizes $L=84$, 100, and 150. As the exponential fit shows (dotted lines), the correlation function exponentially decays to a constant different than zero, with correlation length increasing with system size. These results imply that superfluidity, though suppressed, persists for large values of site dilution even at finite $V/J$. This can be understood using similar arguments as for the case of $V/J=0$. 
As already observed for $V/J=0$ and shown in the bottom-left insert of Figure~\ref{FIG5}, compressibility increases at large enough values of site dilution. This increase is less dramatic than for $V/J=0$ due to a finite value of interaction which suppresses density fluctuations.


\section{Experimental realization}

The model studied in this manuscript has been realized by loading KRb polar molecules into a deep optical lattice where hopping is suppressed. Molecules are cooled so that their hyperfine structures can be ignored. A spin $1/2$ is encoded in the rotational degrees of freedom and microwave fields are used to induce transitions between two different rotational states.
Spin interactions are directly driven by dipolar interactions without the requirement of particle tunneling. 
The ratio $J_\perp/J_z$ can be controlled by tuning the external DC electric field as well as by changing the choice of the rotational states which form the effective spin $1/2$ system~\cite{Wall:2014kw, Gadway:2016er}. 
For example, for rotational states used in~\cite{Yan:2013fn}, $J_z/J_\perp$ can be tuned from 0 to 3 using $\mathbf{E}$ fields from 0 to 16 kV/cm. Based on our results, for this experimental setup, only the easy-plane ferromagnetic order (equivalent to a superfluid phase of model~(\ref{eq:H}) ) may be observed. By choosing different rotational states~\cite{Hazzard:2013hk}, $J_z/J_\perp>10$ can be achieved with similar $\mathbf{E}$ fields, so that easy-axis antiferromagnetic order (equivalent to checkerboard order) and the supersolid phase (where both easy-plane ferromagnetic and easy-axis antiferromagnetic order coexist) may be observed.  

As for the site-diluted case, at the moment, filling factors achieved experimentally are $\lesssim 25\%$ corresponding to a site-dilution $p\gtrsim 0.75 $ in model~\ref{eq:H}. At such large site-dilution, even for $J_z/J_\perp=0$, the easy-plane ferromagnetic order is strongly suppressed. We have estimated that temperatures $T\sim1.5$ nK are needed in order to observe easy-plane ferromegnetic order. Larger filling factors, corresponding to smaller site-dilution, are a little more favorable for observing this phase. We have estimated that with site dilution $p=0.5$, the easy-plane ferromagnetic order would be observed at temperature $T\sim 6.0$ nK. 

\section{Conclusion}

Recent experiments with ultracold polar molecules trapped in deep optical lattices, where a spin 1/2 degree of freedom is encoded in two internal states, realize the site-diluted long-ranged XXZ model. Site dilution results in off-diagonal disorder. We have mapped the XXZ model onto the hard-core Bose-Hubbard model and have performed large-scale Monte Carlo simulations by the Worm algorithm. We have studied the ground state phase diagram of the model in the absence of site-dilution. We have found that, 
for large enough interaction, three phases are stabilized: a superfluid phase, a checkerboard solid phase, only present at density $m=0.5$, and a checkerboard supersolid phase which can be reached by doping the CB phase away from half-filling. We have also studied the model in the presence of site-dilution and at fixed density $m=0.5$. We have found that, unlike what observed in the case of short-range hopping, localization never occurs even for site dilution larger than the percolation threshold and off-diagonal order, though strongly suppressed, persists for arbitrarily large values of site-dilution. We have notice that, a suppressed SF density will result in a suppressed critical temperature for the SF to normal liquid transition at finite temperature, and have provided estimates for temperatures needed to observe coherence. Interestingly, we have found that compressibility increases as site dilution increases. \\
{\textit{Acknowledgements}}  We would like to thank A. Kuklov and A. Safavi-Naini for enlightening discussions. This work was supported by the NSF (PIF-1552978). The computing for this project was performed at the OU Supercomputing Center for Education and Research (OSCER) at the University of Oklahoma (OU).

\bibliography{longv1}

\begin{thebibliography}{35}%
\makeatletter
\providecommand \@ifxundefined [1]{%
 \@ifx{#1\undefined}
}%
\providecommand \@ifnum [1]{%
 \ifnum #1\expandafter \@firstoftwo
 \else \expandafter \@secondoftwo
 \fi
}%
\providecommand \@ifx [1]{%
 \ifx #1\expandafter \@firstoftwo
 \else \expandafter \@secondoftwo
 \fi
}%
\providecommand \natexlab [1]{#1}%
\providecommand \enquote  [1]{``#1''}%
\providecommand \bibnamefont  [1]{#1}%
\providecommand \bibfnamefont [1]{#1}%
\providecommand \citenamefont [1]{#1}%
\providecommand \href@noop [0]{\@secondoftwo}%
\providecommand \href [0]{\begingroup \@sanitize@url \@href}%
\providecommand \@href[1]{\@@startlink{#1}\@@href}%
\providecommand \@@href[1]{\endgroup#1\@@endlink}%
\providecommand \@sanitize@url [0]{\catcode `\\12\catcode `\$12\catcode
  `\&12\catcode `\#12\catcode `\^12\catcode `\_12\catcode `\%12\relax}%
\providecommand \@@startlink[1]{}%
\providecommand \@@endlink[0]{}%
\providecommand \url  [0]{\begingroup\@sanitize@url \@url }%
\providecommand \@url [1]{\endgroup\@href {#1}{\urlprefix }}%
\providecommand \urlprefix  [0]{URL }%
\providecommand \Eprint [0]{\href }%
\providecommand \doibase [0]{http://dx.doi.org/}%
\providecommand \selectlanguage [0]{\@gobble}%
\providecommand \bibinfo  [0]{\@secondoftwo}%
\providecommand \bibfield  [0]{\@secondoftwo}%
\providecommand \translation [1]{[#1]}%
\providecommand \BibitemOpen [0]{}%
\providecommand \bibitemStop [0]{}%
\providecommand \bibitemNoStop [0]{.\EOS\space}%
\providecommand \EOS [0]{\spacefactor3000\relax}%
\providecommand \BibitemShut  [1]{\csname bibitem#1\endcsname}%
\let\auto@bib@innerbib\@empty
\bibitem [{\citenamefont {Crowell}\ \emph {et~al.}(1995)\citenamefont
  {Crowell}, \citenamefont {Van~Keuls},\ and\ \citenamefont
  {Reppy}}]{Crowell:1995cx}%
  \BibitemOpen
  \bibfield  {author} {\bibinfo {author} {\bibfnamefont {P.~A.}\ \bibnamefont
  {Crowell}}, \bibinfo {author} {\bibfnamefont {F.~W.}\ \bibnamefont
  {Van~Keuls}}, \ and\ \bibinfo {author} {\bibfnamefont {J.~D.}\ \bibnamefont
  {Reppy}},\ }\href {\doibase 10.1103/PhysRevLett.75.1106} {\bibfield
  {journal} {\bibinfo  {journal} {Phys. Rev. Lett.}\ }\textbf {\bibinfo
  {volume} {75}},\ \bibinfo {pages} {1106} (\bibinfo {year}
  {1995})}\BibitemShut {NoStop}%
\bibitem [{\citenamefont {Crowell}\ \emph {et~al.}(1997)\citenamefont
  {Crowell}, \citenamefont {Van~Keuls},\ and\ \citenamefont
  {Reppy}}]{Crowell:1997gm}%
  \BibitemOpen
  \bibfield  {author} {\bibinfo {author} {\bibfnamefont {P.~A.}\ \bibnamefont
  {Crowell}}, \bibinfo {author} {\bibfnamefont {F.~W.}\ \bibnamefont
  {Van~Keuls}}, \ and\ \bibinfo {author} {\bibfnamefont {J.~D.}\ \bibnamefont
  {Reppy}},\ }\href {\doibase 10.1103/PhysRevB.55.12620} {\bibfield  {journal}
  {\bibinfo  {journal} {Phys. Rev. B}\ }\textbf {\bibinfo {volume} {55}},\
  \bibinfo {pages} {12620} (\bibinfo {year} {1997})}\BibitemShut {NoStop}%
\bibitem [{\citenamefont {Reppy}\ \emph {et~al.}(2000)\citenamefont {Reppy},
  \citenamefont {Crooker}, \citenamefont {Hebral}, \citenamefont {Corwin},
  \citenamefont {He},\ and\ \citenamefont {Zassenhaus}}]{Reppy:2000cb}%
  \BibitemOpen
  \bibfield  {author} {\bibinfo {author} {\bibfnamefont {J.~D.}\ \bibnamefont
  {Reppy}}, \bibinfo {author} {\bibfnamefont {B.~C.}\ \bibnamefont {Crooker}},
  \bibinfo {author} {\bibfnamefont {B.}~\bibnamefont {Hebral}}, \bibinfo
  {author} {\bibfnamefont {A.~D.}\ \bibnamefont {Corwin}}, \bibinfo {author}
  {\bibfnamefont {J.}~\bibnamefont {He}}, \ and\ \bibinfo {author}
  {\bibfnamefont {G.~M.}\ \bibnamefont {Zassenhaus}},\ }\href {\doibase
  10.1103/PhysRevLett.84.2060} {\bibfield  {journal} {\bibinfo  {journal}
  {Phys. Rev. Lett.}\ }\textbf {\bibinfo {volume} {84}},\ \bibinfo {pages}
  {2060} (\bibinfo {year} {2000})}\BibitemShut {NoStop}%
\bibitem [{\citenamefont {Goldman}\ and\ \citenamefont
  {Markovi{\'c}}(1998)}]{Goldman:1998jy}%
  \BibitemOpen
  \bibfield  {author} {\bibinfo {author} {\bibfnamefont {A.~M.}\ \bibnamefont
  {Goldman}}\ and\ \bibinfo {author} {\bibfnamefont {N.}~\bibnamefont
  {Markovi{\'c}}},\ }\href {\doibase 10.1063/1.882069} {\bibfield  {journal}
  {\bibinfo  {journal} {Phys. Today}\ }\textbf {\bibinfo {volume} {51}},\
  \bibinfo {pages} {39} (\bibinfo {year} {1998})}\BibitemShut {NoStop}%
\bibitem [{\citenamefont {Vitkalov}\ \emph {et~al.}(2001)\citenamefont
  {Vitkalov}, \citenamefont {Zheng}, \citenamefont {Mertes}, \citenamefont
  {Sarachik},\ and\ \citenamefont {Klapwijk}}]{Vitkalov:2001in}%
  \BibitemOpen
  \bibfield  {author} {\bibinfo {author} {\bibfnamefont {S.~A.}\ \bibnamefont
  {Vitkalov}}, \bibinfo {author} {\bibfnamefont {H.}~\bibnamefont {Zheng}},
  \bibinfo {author} {\bibfnamefont {K.~M.}\ \bibnamefont {Mertes}}, \bibinfo
  {author} {\bibfnamefont {M.~P.}\ \bibnamefont {Sarachik}}, \ and\ \bibinfo
  {author} {\bibfnamefont {T.~M.}\ \bibnamefont {Klapwijk}},\ }\href {\doibase
  10.1103/PhysRevLett.87.086401} {\bibfield  {journal} {\bibinfo  {journal}
  {Phys. Rev. Lett.}\ }\textbf {\bibinfo {volume} {87}},\ \bibinfo {pages}
  {251} (\bibinfo {year} {2001})}\BibitemShut {NoStop}%
\bibitem [{\citenamefont {Schneider}\ \emph {et~al.}(2012)\citenamefont
  {Schneider}, \citenamefont {Zaitsev}, \citenamefont {Fuchs},\ and\
  \citenamefont {von L{\"o}hneysen}}]{Schneider:2012id}%
  \BibitemOpen
  \bibfield  {author} {\bibinfo {author} {\bibfnamefont {R.}~\bibnamefont
  {Schneider}}, \bibinfo {author} {\bibfnamefont {A.~G.}\ \bibnamefont
  {Zaitsev}}, \bibinfo {author} {\bibfnamefont {D.}~\bibnamefont {Fuchs}}, \
  and\ \bibinfo {author} {\bibfnamefont {H.}~\bibnamefont {von
  L{\"o}hneysen}},\ }\href {\doibase 10.1103/PhysRevLett.108.257003} {\bibfield
   {journal} {\bibinfo  {journal} {Phys. Rev. Lett.}\ }\textbf {\bibinfo
  {volume} {108}},\ \bibinfo {pages} {257003} (\bibinfo {year}
  {2012})}\BibitemShut {NoStop}%
\bibitem [{\citenamefont {van~der Zant}\ \emph {et~al.}(1996)\citenamefont
  {van~der Zant}, \citenamefont {Elion}, \citenamefont {Geerligs},\ and\
  \citenamefont {Mooij}}]{vanderZant:1996cf}%
  \BibitemOpen
  \bibfield  {author} {\bibinfo {author} {\bibfnamefont {H.~S.~J.}\
  \bibnamefont {van~der Zant}}, \bibinfo {author} {\bibfnamefont {W.~J.}\
  \bibnamefont {Elion}}, \bibinfo {author} {\bibfnamefont {L.~J.}\ \bibnamefont
  {Geerligs}}, \ and\ \bibinfo {author} {\bibfnamefont {J.~E.}\ \bibnamefont
  {Mooij}},\ }\href {\doibase 10.1103/PhysRevB.54.10081} {\bibfield  {journal}
  {\bibinfo  {journal} {Phys. Rev. B}\ }\textbf {\bibinfo {volume} {54}},\
  \bibinfo {pages} {10081} (\bibinfo {year} {1996})}\BibitemShut {NoStop}%
\bibitem [{\citenamefont {White}\ \emph {et~al.}(2009)\citenamefont {White},
  \citenamefont {Pasienski}, \citenamefont {McKay}, \citenamefont {Zhou},
  \citenamefont {Ceperley},\ and\ \citenamefont {DeMarco}}]{DeMarco2009PRL}%
  \BibitemOpen
  \bibfield  {author} {\bibinfo {author} {\bibfnamefont {M.}~\bibnamefont
  {White}}, \bibinfo {author} {\bibfnamefont {M.}~\bibnamefont {Pasienski}},
  \bibinfo {author} {\bibfnamefont {D.}~\bibnamefont {McKay}}, \bibinfo
  {author} {\bibfnamefont {S.~Q.}\ \bibnamefont {Zhou}}, \bibinfo {author}
  {\bibfnamefont {D.}~\bibnamefont {Ceperley}}, \ and\ \bibinfo {author}
  {\bibfnamefont {B.}~\bibnamefont {DeMarco}},\ }\href {\doibase
  10.1103/PhysRevLett.102.055301} {\bibfield  {journal} {\bibinfo  {journal}
  {Phys. Rev. Lett.}\ }\textbf {\bibinfo {volume} {102}},\ \bibinfo {pages}
  {055301} (\bibinfo {year} {2009})}\BibitemShut {NoStop}%
\bibitem [{\citenamefont {Pasienski}\ \emph {et~al.}(2010)\citenamefont
  {Pasienski}, \citenamefont {McKay}, \citenamefont {White},\ and\
  \citenamefont {DeMarco}}]{DeMarco2010Nature}%
  \BibitemOpen
  \bibfield  {author} {\bibinfo {author} {\bibfnamefont {M.}~\bibnamefont
  {Pasienski}}, \bibinfo {author} {\bibfnamefont {D.}~\bibnamefont {McKay}},
  \bibinfo {author} {\bibfnamefont {M.}~\bibnamefont {White}}, \ and\ \bibinfo
  {author} {\bibfnamefont {B.}~\bibnamefont {DeMarco}},\ }\href {\doibase
  10.1038/nphys1726} {\bibfield  {journal} {\bibinfo  {journal} {Nat. Phys.}\
  }\textbf {\bibinfo {volume} {6}},\ \bibinfo {pages} {677} (\bibinfo {year}
  {2010})}\BibitemShut {NoStop}%
\bibitem [{\citenamefont {Roati}\ \emph {et~al.}(2008)\citenamefont {Roati},
  \citenamefont {D'Errico}, \citenamefont {Fallani}, \citenamefont {Fattori},
  \citenamefont {Fort}, \citenamefont {Zaccanti}, \citenamefont {Modugno},
  \citenamefont {Modugno},\ and\ \citenamefont {Inguscio}}]{Inguscio2008}%
  \BibitemOpen
  \bibfield  {author} {\bibinfo {author} {\bibfnamefont {G.}~\bibnamefont
  {Roati}}, \bibinfo {author} {\bibfnamefont {C.}~\bibnamefont {D'Errico}},
  \bibinfo {author} {\bibfnamefont {L.}~\bibnamefont {Fallani}}, \bibinfo
  {author} {\bibfnamefont {M.}~\bibnamefont {Fattori}}, \bibinfo {author}
  {\bibfnamefont {C.}~\bibnamefont {Fort}}, \bibinfo {author} {\bibfnamefont
  {M.}~\bibnamefont {Zaccanti}}, \bibinfo {author} {\bibfnamefont
  {G.}~\bibnamefont {Modugno}}, \bibinfo {author} {\bibfnamefont
  {M.}~\bibnamefont {Modugno}}, \ and\ \bibinfo {author} {\bibfnamefont
  {M.}~\bibnamefont {Inguscio}},\ }\href {\doibase 10.1038/nature07071}
  {\bibfield  {journal} {\bibinfo  {journal} {Nature}\ }\textbf {\bibinfo
  {volume} {453}},\ \bibinfo {pages} {895} (\bibinfo {year}
  {2008})}\BibitemShut {NoStop}%
\bibitem [{\citenamefont {Rapsch}\ \emph {et~al.}(1999)\citenamefont {Rapsch},
  \citenamefont {Schollwock},\ and\ \citenamefont {Zwerger}}]{Rapsch:1999bw}%
  \BibitemOpen
  \bibfield  {author} {\bibinfo {author} {\bibfnamefont {S.}~\bibnamefont
  {Rapsch}}, \bibinfo {author} {\bibfnamefont {U.}~\bibnamefont {Schollwock}},
  \ and\ \bibinfo {author} {\bibfnamefont {W.}~\bibnamefont {Zwerger}},\ }\href
  {\doibase 10.1209/epl/i1999-00302-7} {\bibfield  {journal} {\bibinfo
  {journal} {Europhy. Lett.}\ }\textbf {\bibinfo {volume} {46}},\ \bibinfo
  {pages} {559} (\bibinfo {year} {1999})}\BibitemShut {NoStop}%
\bibitem [{\citenamefont {Krauth}\ \emph {et~al.}(1991)\citenamefont {Krauth},
  \citenamefont {Trivedi},\ and\ \citenamefont {Ceperley}}]{Ceperley1991}%
  \BibitemOpen
  \bibfield  {author} {\bibinfo {author} {\bibfnamefont {W.}~\bibnamefont
  {Krauth}}, \bibinfo {author} {\bibfnamefont {N.}~\bibnamefont {Trivedi}}, \
  and\ \bibinfo {author} {\bibfnamefont {D.}~\bibnamefont {Ceperley}},\ }\href
  {http://journals.aps.org/prl/abstract/10.1103/PhysRevLett.67.2307} {\bibfield
   {journal} {\bibinfo  {journal} {Phys. Rev. Lett.}\ }\textbf {\bibinfo
  {volume} {67}},\ \bibinfo {pages} {2307} (\bibinfo {year}
  {1991})}\BibitemShut {NoStop}%
\bibitem [{\citenamefont {Pollet}\ \emph {et~al.}(2009)\citenamefont {Pollet},
  \citenamefont {Prokof'ev}, \citenamefont {Svistunov},\ and\ \citenamefont
  {Troyer}}]{Pollet2009BG}%
  \BibitemOpen
  \bibfield  {author} {\bibinfo {author} {\bibfnamefont {L.}~\bibnamefont
  {Pollet}}, \bibinfo {author} {\bibfnamefont {N.}~\bibnamefont {Prokof'ev}},
  \bibinfo {author} {\bibfnamefont {B.}~\bibnamefont {Svistunov}}, \ and\
  \bibinfo {author} {\bibfnamefont {M.}~\bibnamefont {Troyer}},\ }\href
  {\doibase 10.1103/PhysRevLett.103.140402} {\bibfield  {journal} {\bibinfo
  {journal} {Phys. Rev. Lett.}\ }\textbf {\bibinfo {volume} {103}},\ \bibinfo
  {pages} {140402} (\bibinfo {year} {2009})}\BibitemShut {NoStop}%
\bibitem [{\citenamefont {Gurarie}\ \emph {et~al.}(2009)\citenamefont
  {Gurarie}, \citenamefont {Pollet}, \citenamefont {Prokof'ev}, \citenamefont
  {Svistunov},\ and\ \citenamefont {Troyer}}]{Pollet2009PRB}%
  \BibitemOpen
  \bibfield  {author} {\bibinfo {author} {\bibfnamefont {V.}~\bibnamefont
  {Gurarie}}, \bibinfo {author} {\bibfnamefont {L.}~\bibnamefont {Pollet}},
  \bibinfo {author} {\bibfnamefont {N.~V.}\ \bibnamefont {Prokof'ev}}, \bibinfo
  {author} {\bibfnamefont {B.~V.}\ \bibnamefont {Svistunov}}, \ and\ \bibinfo
  {author} {\bibfnamefont {M.}~\bibnamefont {Troyer}},\ }\href {\doibase
  10.1103/PhysRevB.80.214519} {\bibfield  {journal} {\bibinfo  {journal} {Phys.
  Rev. B}\ }\textbf {\bibinfo {volume} {80}},\ \bibinfo {pages} {214519}
  (\bibinfo {year} {2009})}\BibitemShut {NoStop}%
\bibitem [{\citenamefont {Soyler}\ \emph {et~al.}(2011)\citenamefont {Soyler},
  \citenamefont {Kiselev}, \citenamefont {Prokof'ev},\ and\ \citenamefont
  {Svistunov}}]{Soyler:2011ik}%
  \BibitemOpen
  \bibfield  {author} {\bibinfo {author} {\bibfnamefont {S.~G.}\ \bibnamefont
  {Soyler}}, \bibinfo {author} {\bibfnamefont {M.}~\bibnamefont {Kiselev}},
  \bibinfo {author} {\bibfnamefont {N.~V.}\ \bibnamefont {Prokof'ev}}, \ and\
  \bibinfo {author} {\bibfnamefont {B.~V.}\ \bibnamefont {Svistunov}},\ }\href
  {\doibase 10.1103/PhysRevLett.107.185301} {\bibfield  {journal} {\bibinfo
  {journal} {Phys. Rev. Lett.}\ }\textbf {\bibinfo {volume} {107}},\ \bibinfo
  {pages} {185301} (\bibinfo {year} {2011})}\BibitemShut {NoStop}%
\bibitem [{\citenamefont {Lin}\ \emph {et~al.}(2011)\citenamefont {Lin},
  \citenamefont {S{\o}rensen},\ and\ \citenamefont {Ceperley}}]{Ceperley2011}%
  \BibitemOpen
  \bibfield  {author} {\bibinfo {author} {\bibfnamefont {F.}~\bibnamefont
  {Lin}}, \bibinfo {author} {\bibfnamefont {E.~S.}\ \bibnamefont
  {S{\o}rensen}}, \ and\ \bibinfo {author} {\bibfnamefont {D.~M.}\ \bibnamefont
  {Ceperley}},\ }\href
  {http://journals.aps.org/prb/abstract/10.1103/PhysRevB.84.094507} {\bibfield
  {journal} {\bibinfo  {journal} {Phys. Rev. B}\ }\textbf {\bibinfo {volume}
  {84}},\ \bibinfo {pages} {094507} (\bibinfo {year} {2011})}\BibitemShut
  {NoStop}%
\bibitem [{\citenamefont {Zhang}\ \emph {et~al.}(2015)\citenamefont {Zhang},
  \citenamefont {Safavi-Naini},\ and\ \citenamefont
  {Capogrosso-Sansone}}]{Zhang:2015it}%
  \BibitemOpen
  \bibfield  {author} {\bibinfo {author} {\bibfnamefont {C.}~\bibnamefont
  {Zhang}}, \bibinfo {author} {\bibfnamefont {A.}~\bibnamefont {Safavi-Naini}},
  \ and\ \bibinfo {author} {\bibfnamefont {B.}~\bibnamefont
  {Capogrosso-Sansone}},\ }\href {\doibase 10.1103/PhysRevA.91.031604}
  {\bibfield  {journal} {\bibinfo  {journal} {Phys. Rev. A}\ }\textbf {\bibinfo
  {volume} {91}},\ \bibinfo {pages} {031604} (\bibinfo {year}
  {2015})}\BibitemShut {NoStop}%
\bibitem [{\citenamefont {Niederle}\ and\ \citenamefont
  {Rieger}(2013)}]{Niederle:2013jy}%
  \BibitemOpen
  \bibfield  {author} {\bibinfo {author} {\bibfnamefont {A.~E.}\ \bibnamefont
  {Niederle}}\ and\ \bibinfo {author} {\bibfnamefont {H.}~\bibnamefont
  {Rieger}},\ }\href {\doibase 10.1088/1367-2630/15/7/075029} {\bibfield
  {journal} {\bibinfo  {journal} {New J Phys}\ }\textbf {\bibinfo {volume}
  {15}},\ \bibinfo {pages} {075029} (\bibinfo {year} {2013})}\BibitemShut
  {NoStop}%
\bibitem [{\citenamefont {Levitov}(1990)}]{Levitov:1990jd}%
  \BibitemOpen
  \bibfield  {author} {\bibinfo {author} {\bibfnamefont {L.~S.}\ \bibnamefont
  {Levitov}},\ }\href {\doibase 10.1103/PhysRevLett.64.547} {\bibfield
  {journal} {\bibinfo  {journal} {Phys. Rev. Lett.}\ }\textbf {\bibinfo
  {volume} {64}},\ \bibinfo {pages} {547} (\bibinfo {year} {1990})}\BibitemShut
  {NoStop}%
\bibitem [{\citenamefont {Sengupta}\ and\ \citenamefont
  {Haas}(2007)}]{Sengupta:2007jx}%
  \BibitemOpen
  \bibfield  {author} {\bibinfo {author} {\bibfnamefont {P.}~\bibnamefont
  {Sengupta}}\ and\ \bibinfo {author} {\bibfnamefont {S.}~\bibnamefont
  {Haas}},\ }\href {\doibase 10.1103/PhysRevLett.99.050403} {\bibfield
  {journal} {\bibinfo  {journal} {Phys. Rev. Lett.}\ }\textbf {\bibinfo
  {volume} {99}},\ \bibinfo {pages} {050403} (\bibinfo {year}
  {2007})}\BibitemShut {NoStop}%
\bibitem [{\citenamefont {Prokof'ev}\ and\ \citenamefont
  {Svistunov}(2004)}]{Prokofev:2004fk}%
  \BibitemOpen
  \bibfield  {author} {\bibinfo {author} {\bibfnamefont {N.}~\bibnamefont
  {Prokof'ev}}\ and\ \bibinfo {author} {\bibfnamefont {B.}~\bibnamefont
  {Svistunov}},\ }\href {\doibase 10.1103/PhysRevLett.92.015703} {\bibfield
  {journal} {\bibinfo  {journal} {Phys. Rev. Lett.}\ }\textbf {\bibinfo
  {volume} {92}},\ \bibinfo {pages} {015703} (\bibinfo {year}
  {2004})}\BibitemShut {NoStop}%
\bibitem [{\citenamefont {Roscilde}\ and\ \citenamefont
  {Haas}(2007)}]{Roscilde:2007ch}%
  \BibitemOpen
  \bibfield  {author} {\bibinfo {author} {\bibfnamefont {T.}~\bibnamefont
  {Roscilde}}\ and\ \bibinfo {author} {\bibfnamefont {S.}~\bibnamefont
  {Haas}},\ }\href {\doibase 10.1103/PhysRevLett.99.047205} {\bibfield
  {journal} {\bibinfo  {journal} {Phys. Rev. Lett.}\ }\textbf {\bibinfo
  {volume} {99}},\ \bibinfo {pages} {047205} (\bibinfo {year}
  {2007})}\BibitemShut {NoStop}%
\bibitem [{\citenamefont {Vojta}\ \emph {et~al.}(2016)\citenamefont {Vojta},
  \citenamefont {Crewse}, \citenamefont {Puschmann}, \citenamefont {Arovas},\
  and\ \citenamefont {Kiselev}}]{Vojta:2016kl}%
  \BibitemOpen
  \bibfield  {author} {\bibinfo {author} {\bibfnamefont {T.}~\bibnamefont
  {Vojta}}, \bibinfo {author} {\bibfnamefont {J.}~\bibnamefont {Crewse}},
  \bibinfo {author} {\bibfnamefont {M.}~\bibnamefont {Puschmann}}, \bibinfo
  {author} {\bibfnamefont {D.}~\bibnamefont {Arovas}}, \ and\ \bibinfo {author}
  {\bibfnamefont {Y.}~\bibnamefont {Kiselev}},\ }\href {\doibase
  10.1103/PhysRevB.94.134501} {\bibfield  {journal} {\bibinfo  {journal} {Phys.
  Rev. B}\ }\textbf {\bibinfo {volume} {94}},\ \bibinfo {pages} {134501}
  (\bibinfo {year} {2016})}\BibitemShut {NoStop}%
\bibitem [{\citenamefont {Yan}\ \emph {et~al.}(2013)\citenamefont {Yan},
  \citenamefont {Moses}, \citenamefont {Gadway}, \citenamefont {Covey},
  \citenamefont {Hazzard}, \citenamefont {Rey}, \citenamefont {Jin},\ and\
  \citenamefont {Ye}}]{Yan:2013fn}%
  \BibitemOpen
  \bibfield  {author} {\bibinfo {author} {\bibfnamefont {B.}~\bibnamefont
  {Yan}}, \bibinfo {author} {\bibfnamefont {S.~A.}\ \bibnamefont {Moses}},
  \bibinfo {author} {\bibfnamefont {B.}~\bibnamefont {Gadway}}, \bibinfo
  {author} {\bibfnamefont {J.~P.}\ \bibnamefont {Covey}}, \bibinfo {author}
  {\bibfnamefont {K.~R.~A.}\ \bibnamefont {Hazzard}}, \bibinfo {author}
  {\bibfnamefont {A.~M.}\ \bibnamefont {Rey}}, \bibinfo {author} {\bibfnamefont
  {D.~S.}\ \bibnamefont {Jin}}, \ and\ \bibinfo {author} {\bibfnamefont
  {J.}~\bibnamefont {Ye}},\ }\href {\doibase 10.1038/nature12483} {\bibfield
  {journal} {\bibinfo  {journal} {Nature}\ }\textbf {\bibinfo {volume} {501}},\
  \bibinfo {pages} {521} (\bibinfo {year} {2013})}\BibitemShut {NoStop}%
\bibitem [{\citenamefont {Hazzard}\ \emph {et~al.}(2014)\citenamefont
  {Hazzard}, \citenamefont {Gadway}, \citenamefont {Foss-Feig}, \citenamefont
  {Yan}, \citenamefont {Moses}, \citenamefont {Covey}, \citenamefont {Yao},
  \citenamefont {Lukin}, \citenamefont {Ye}, \citenamefont {Jin},\ and\
  \citenamefont {Rey}}]{Hazzard:2014bx}%
  \BibitemOpen
  \bibfield  {author} {\bibinfo {author} {\bibfnamefont {K.~R.~A.}\
  \bibnamefont {Hazzard}}, \bibinfo {author} {\bibfnamefont {B.}~\bibnamefont
  {Gadway}}, \bibinfo {author} {\bibfnamefont {M.}~\bibnamefont {Foss-Feig}},
  \bibinfo {author} {\bibfnamefont {B.}~\bibnamefont {Yan}}, \bibinfo {author}
  {\bibfnamefont {S.~A.}\ \bibnamefont {Moses}}, \bibinfo {author}
  {\bibfnamefont {J.~P.}\ \bibnamefont {Covey}}, \bibinfo {author}
  {\bibfnamefont {N.~Y.}\ \bibnamefont {Yao}}, \bibinfo {author} {\bibfnamefont
  {M.~D.}\ \bibnamefont {Lukin}}, \bibinfo {author} {\bibfnamefont
  {J.}~\bibnamefont {Ye}}, \bibinfo {author} {\bibfnamefont {D.~S.}\
  \bibnamefont {Jin}}, \ and\ \bibinfo {author} {\bibfnamefont {A.~M.}\
  \bibnamefont {Rey}},\ }\href {\doibase 10.1103/PhysRevLett.113.195302}
  {\bibfield  {journal} {\bibinfo  {journal} {Phys. Rev. Lett.}\ }\textbf
  {\bibinfo {volume} {113}},\ \bibinfo {pages} {195302} (\bibinfo {year}
  {2014})}\BibitemShut {NoStop}%
\bibitem [{\citenamefont {Moses}\ \emph {et~al.}(2015)\citenamefont {Moses},
  \citenamefont {Covey}, \citenamefont {Miecnikowski}, \citenamefont {Yan},
  \citenamefont {Gadway}, \citenamefont {Ye},\ and\ \citenamefont
  {Jin}}]{Moses:2015df}%
  \BibitemOpen
  \bibfield  {author} {\bibinfo {author} {\bibfnamefont {S.~A.}\ \bibnamefont
  {Moses}}, \bibinfo {author} {\bibfnamefont {J.~P.}\ \bibnamefont {Covey}},
  \bibinfo {author} {\bibfnamefont {M.~T.}\ \bibnamefont {Miecnikowski}},
  \bibinfo {author} {\bibfnamefont {B.}~\bibnamefont {Yan}}, \bibinfo {author}
  {\bibfnamefont {B.}~\bibnamefont {Gadway}}, \bibinfo {author} {\bibfnamefont
  {J.}~\bibnamefont {Ye}}, \ and\ \bibinfo {author} {\bibfnamefont {D.~S.}\
  \bibnamefont {Jin}},\ }\href {\doibase 10.1126/science.aac6400} {\bibfield
  {journal} {\bibinfo  {journal} {Science}\ }\textbf {\bibinfo {volume}
  {350}},\ \bibinfo {pages} {659} (\bibinfo {year} {2015})}\BibitemShut
  {NoStop}%
\bibitem [{\citenamefont {Deng}\ \emph {et~al.}(2017)\citenamefont {Deng},
  \citenamefont {Kravtsov}, \citenamefont {Shlyapnikov},\ and\ \citenamefont
  {Santos}}]{Deng:2017uu}%
  \BibitemOpen
  \bibfield  {author} {\bibinfo {author} {\bibfnamefont {X.}~\bibnamefont
  {Deng}}, \bibinfo {author} {\bibfnamefont {V.~E.}\ \bibnamefont {Kravtsov}},
  \bibinfo {author} {\bibfnamefont {G.~V.}\ \bibnamefont {Shlyapnikov}}, \ and\
  \bibinfo {author} {\bibfnamefont {L.}~\bibnamefont {Santos}},\ }\href
  {http://arxiv.org/abs/1706.04088v1} {\bibfield  {journal} {\bibinfo
  {journal} {arXiv: 1504.04578}\ } (\bibinfo {year} {2017})}\BibitemShut
  {NoStop}%
\bibitem [{\citenamefont {Deng}\ \emph {et~al.}(2016)\citenamefont {Deng},
  \citenamefont {Altshuler}, \citenamefont {Shlyapnikov},\ and\ \citenamefont
  {Santos}}]{Deng:2016gs}%
  \BibitemOpen
  \bibfield  {author} {\bibinfo {author} {\bibfnamefont {X.}~\bibnamefont
  {Deng}}, \bibinfo {author} {\bibfnamefont {B.~L.}\ \bibnamefont {Altshuler}},
  \bibinfo {author} {\bibfnamefont {G.~V.}\ \bibnamefont {Shlyapnikov}}, \ and\
  \bibinfo {author} {\bibfnamefont {L.}~\bibnamefont {Santos}},\ }\href
  {\doibase 10.1103/PhysRevLett.117.020401} {\bibfield  {journal} {\bibinfo
  {journal} {Phys. Rev. Lett.}\ }\textbf {\bibinfo {volume} {117}},\ \bibinfo
  {pages} {441} (\bibinfo {year} {2016})}\BibitemShut {NoStop}%
\bibitem [{\citenamefont {Prokof'ev}\ \emph {et~al.}(1998)\citenamefont
  {Prokof'ev}, \citenamefont {Svistunov},\ and\ \citenamefont
  {Tupitsyn}}]{Prokofev:1998gz}%
  \BibitemOpen
  \bibfield  {author} {\bibinfo {author} {\bibfnamefont {N.~V.}\ \bibnamefont
  {Prokof'ev}}, \bibinfo {author} {\bibfnamefont {B.~V.}\ \bibnamefont
  {Svistunov}}, \ and\ \bibinfo {author} {\bibfnamefont {I.~S.}\ \bibnamefont
  {Tupitsyn}},\ }\href {\doibase 10.1134/1.558661} {\bibfield  {journal}
  {\bibinfo  {journal} {J. Exp. Theor. Phys.}\ }\textbf {\bibinfo {volume}
  {87}},\ \bibinfo {pages} {310} (\bibinfo {year} {1998})}\BibitemShut
  {NoStop}%
\bibitem [{\citenamefont {Pollock}\ and\ \citenamefont
  {Ceperley}(1987)}]{Winding}%
  \BibitemOpen
  \bibfield  {author} {\bibinfo {author} {\bibfnamefont {E.~L.}\ \bibnamefont
  {Pollock}}\ and\ \bibinfo {author} {\bibfnamefont {D.~M.}\ \bibnamefont
  {Ceperley}},\ }\href {\doibase 10.1103/PhysRevB.36.8343} {\bibfield
  {journal} {\bibinfo  {journal} {Phys. Rev. B}\ }\textbf {\bibinfo {volume}
  {36}},\ \bibinfo {pages} {8343} (\bibinfo {year} {1987})}\BibitemShut
  {NoStop}%
\bibitem [{\citenamefont {Rousseau}(2014)}]{Rousseau:2014jt}%
  \BibitemOpen
  \bibfield  {author} {\bibinfo {author} {\bibfnamefont {V.~G.}\ \bibnamefont
  {Rousseau}},\ }\href {\doibase 10.1103/PhysRevB.90.134503} {\bibfield
  {journal} {\bibinfo  {journal} {Phys. Rev. B}\ }\textbf {\bibinfo {volume}
  {90}},\ \bibinfo {pages} {134503} (\bibinfo {year} {2014})}\BibitemShut
  {NoStop}%
\bibitem [{\citenamefont {Capogrosso-Sansone}\ \emph
  {et~al.}(2010)\citenamefont {Capogrosso-Sansone}, \citenamefont {Trefzger},
  \citenamefont {Lewenstein}, \citenamefont {Zoller},\ and\ \citenamefont
  {Pupillo}}]{CapogrossoSansone:2010em}%
  \BibitemOpen
  \bibfield  {author} {\bibinfo {author} {\bibfnamefont {B.}~\bibnamefont
  {Capogrosso-Sansone}}, \bibinfo {author} {\bibfnamefont {C.}~\bibnamefont
  {Trefzger}}, \bibinfo {author} {\bibfnamefont {M.}~\bibnamefont
  {Lewenstein}}, \bibinfo {author} {\bibfnamefont {P.}~\bibnamefont {Zoller}},
  \ and\ \bibinfo {author} {\bibfnamefont {G.}~\bibnamefont {Pupillo}},\ }\href
  {\doibase 10.1103/PhysRevLett.104.125301} {\bibfield  {journal} {\bibinfo
  {journal} {Phys. Rev. Lett.}\ }\textbf {\bibinfo {volume} {104}},\ \bibinfo
  {pages} {125301} (\bibinfo {year} {2010})}\BibitemShut {NoStop}%
\bibitem [{\citenamefont {Wall}\ \emph {et~al.}(2014)\citenamefont {Wall},
  \citenamefont {Hazzard},\ and\ \citenamefont {Rey}}]{Wall:2014kw}%
  \BibitemOpen
  \bibfield  {author} {\bibinfo {author} {\bibfnamefont {M.~L.}\ \bibnamefont
  {Wall}}, \bibinfo {author} {\bibfnamefont {K.~R.~A.}\ \bibnamefont
  {Hazzard}}, \ and\ \bibinfo {author} {\bibfnamefont {A.~M.}\ \bibnamefont
  {Rey}},\ }\href {https://arxiv.org/abs/1406.4758} {\bibfield  {journal}
  {\bibinfo  {journal} {arXiv: 1406.4758v1}\ } (\bibinfo {year}
  {2014})}\BibitemShut {NoStop}%
\bibitem [{\citenamefont {Gadway}\ and\ \citenamefont
  {Yan}(2016)}]{Gadway:2016er}%
  \BibitemOpen
  \bibfield  {author} {\bibinfo {author} {\bibfnamefont {B.}~\bibnamefont
  {Gadway}}\ and\ \bibinfo {author} {\bibfnamefont {B.}~\bibnamefont {Yan}},\
  }\href {\doibase 10.1088/0953-4075/49/15/152002} {\bibfield  {journal}
  {\bibinfo  {journal} {J. Phys. B: At. Mol. Opt. Phys.}\ }\textbf {\bibinfo
  {volume} {49}},\ \bibinfo {pages} {152002} (\bibinfo {year}
  {2016})}\BibitemShut {NoStop}%
\bibitem [{\citenamefont {Hazzard}\ \emph {et~al.}(2013)\citenamefont
  {Hazzard}, \citenamefont {Manmana}, \citenamefont {Foss-Feig},\ and\
  \citenamefont {Rey}}]{Hazzard:2013hk}%
  \BibitemOpen
  \bibfield  {author} {\bibinfo {author} {\bibfnamefont {K.~R.~A.}\
  \bibnamefont {Hazzard}}, \bibinfo {author} {\bibfnamefont {S.~R.}\
  \bibnamefont {Manmana}}, \bibinfo {author} {\bibfnamefont {M.}~\bibnamefont
  {Foss-Feig}}, \ and\ \bibinfo {author} {\bibfnamefont {A.~M.}\ \bibnamefont
  {Rey}},\ }\href {\doibase 10.1103/PhysRevLett.110.075301} {\bibfield
  {journal} {\bibinfo  {journal} {Phys. Rev. Lett.}\ }\textbf {\bibinfo
  {volume} {110}},\ \bibinfo {pages} {075301} (\bibinfo {year}
  {2013})}\BibitemShut {NoStop}%
\end{thebibliography}%


\end{document}